# Investigation of passive flow control over an airfoil using leading edge tubercles

Alok Mishra[1], Ashoke De[1,*]

[1]Department of Aerospace Engineering, Indian Institute of Technology Kanpur, Kanpur, 208016, India.

*corresponding author: ashoke@iitk.ac.in

## Abstract

The present paper numerically investigates flow control over NACA0021 (National Advisory Committee for Aeronautics) airfoil by applying a leading-edge tubercle as a passive flow control device at transitional Reynolds number (Re) regime. The study includes simulations over the unmodified airfoil and modified airfoil operating at Re=120000 using $k_T$-$k_L$-ω based Delayed Detached Eddy Simulation (DDES) model. The calculations include both the pre-stall and post-stall regions, while the performance of the mean flow aerodynamic properties like pressure, lift, skin-friction, and drag coefficients are also investigated in the presence of tubercles. The spanwise flow over the modified airfoil interacts with the primary flow. The turbulent separation gets delayed compared to the unmodified airfoil, thereby significantly improving the stall behavior and flow behavior in the post-stall region. The modified airfoil provides the gradual stall behavior against the steep stall behavior for the unmodified airfoil. To optimize tubercle configuration, the study considers two modified airfoils with different combinations of amplitude and wavelength. The unsteady analysis, including the reduced-order modeling, i.e. 3D Proper orthogonal decomposition (POD), characterizes the dominant vortical structure of the flow to evaluate the improvement in the aerodynamic performance of the modified airfoil. The model decomposition for flow over airfoil helps to provide a deeper understanding of the flow control phenomenon.

## I. INTRODUCTION

The flow investigation over an airfoil at a low Reynolds number regime is very significant for military and civil applications such as propeller, micro air vehicle, high altitude sailplane, unmanned aerial vehicles, and wind turbines and turbomachinery blades. At a low Reynolds number, the flow changes from laminar to turbulent flow, and this process is called transition. The crucial fluid flow phenomenon at the low Reynolds number range (Re =60,000 to 200,000) is the laminar separation bubble (LBS). At low Reynolds number flow, the boundary layer over the airfoil or aerodynamic body at the pressure recovery region may still be laminar. The laminar boundary layer cannot withstand the adverse pressure gradient; thus, the flow separates from the surface of the aerodynamic body as a laminar separation. The instability develops in the shear layer of separated flow, which leads to the transition phenomenon at the mid-air, and the laminar shear layer converts into the turbulent shear layer. Once the turbulent shear layer acquires a sufficient amount of energy, it re-attaches as a turbulent boundary layer, and the air in-trap between the laminar boundary layer separation and turbulent boundary layer attachment is called the laminar separation bubble. The laminar separation bubble negatively affects the aerodynamic performance of airfoil and alters stall characteristics. Studying and controlling the laminar separation bubble is essential to improve the overall performance of airfoil at a low Reynolds number regime.[1-9]

Flow control over an aerodynamic body can manipulate the boundary layer to delay the separation from the surface, which, as a result, reduces the pressure drag, increases the lift, and improves overall the performance of the airfoil. The flow control over an airfoil can be divided into two methods, i.e. active flow control and passive flow control: **(a)** If the system is provided with the external energy to control flow, called active flow control and pulsed vortex generator jets (VGJs)[10-12], Plasma actuators[13-17], suction near the trailing edge[18], synthetic jet[19], simultaneous suction and blowing[20-21] and distributed suction over airfoil[22] are some example of the active flow control method. **(b)** If the surface is modified to manipulate the flow over the airfoil, called the passive flow control method. The surface modification can be achieved through altering surface roughness[23], leading-edge protuberances[24], hairy flaps[25], off-surface control element[26], thin control plates[27], and leading-edge slats[28-30]. Because of the high cost and structure complexity, it can be said that active control devices[31] have lagged considerably behind passive flow control devices. The active flow control is very complex and possesses a high operating cost. The passive flow control techniques have a significant advantage over the active flow control techniques due to these drawbacks. Since our work focuses on investigating the passive flow control over an airfoil at a low Reynolds number regime, the rest of the discussion in the introduction section is restricted to the leading-edge tubercle over an airfoil only. The application of leading-edge tubercles on the airfoil as a passive flow control device is inspired by aquatic animal motion. The investigation of the humpback whale's motion reveals that the humpback whale achieves a high degree of maneuverability underwater through the co-ordination of their flapper. The large rounded shape bumps (leading-edge tubercles) along the leading edge of the flipper are unique morphological structures. The tubercles on the leading edge and the flipper of the humpback whale act as passive-flow control devices that improve the whale's performance and maneuverability to capture prey[32-36].

The extensive experimental work was carried out on the humpback whale's flipper to study the effect of leading-edge tubercle on aerodynamic performance. Fish and battle[37] were designed to replicate the flipper of humpback whale into the experimental model to investigate the effect of the rounded bump along the leading edge. The lift of the flipper enhances and maintains at the high angles of attack (AOA) with the tubercle application as a passive flow control device. Miklosovic et al.[38] carried out an experimental study on the stall behavior of the humpback whale flipper applying tubercle on the NACA0021 at Re = 505,000-520,000. The results were compared to the scaled model of a flipper with and without leading-edge tubercle. The study showed a 40% increment in the stall angle for the flipper with the tubercle, which significantly increased the flipper's operating range. An increase of 6% in the total maximum lift coefficient and a decrease in drag up to 32% in the post-stall regime was reported for modification with a rounded bump along the leading edge of the airfoil. The key advantage of leading-edge tubercles is to delay the stall angle and thus provides a significant lift coefficient even at a high angle of incidence. The tubercle at the leading edge, along with the airfoil, acts as a vortex generator device, and these streamwise vortices interacted with flow separation over the airfoil or wing. Hansen et al.[39] experimented with two thick airfoils (NACA-0021 and NACA 65-021) at transition regime (Re = 120,000) with the leading-edge tubercle. The flow over airfoil was investigated for a modified airfoil with different wavelengths ($\lambda$) and amplitude (A). It was reported that a higher maximum lift coefficient and a larger stall angle are associated with the lower tubercle amplitude.

Various researchers reported the advantages of the leading-edge tubercle as available in the open literature[39-42]. Fish et al.[43] suggested the application of bio-inspired technology using leading-edge tubercle. The application of leading-edge tubercles as passive control of flow has been utilized in the design of sailboat centerboards[38], unmanned aerial vehicles[44], boat rudders[45], propellers[46], wind turbines[1], and compressor blade[47].

Watts and Fish[48] conducted Computational Fluid Dynamics (CFD) simulations to study the hydrodynamics of the humpback whale flipper to compare the performance of NACA 63-021 airfoil with leading-edge sinusoidal tubercles with the unmodified airfoil such as 4.8% increase in lift, 10.9% reduction in induced drag, and 17.6% increase in the lift to drag ratio is found at a 10° AOA. Corsini et al.[49] reported a three-dimensional numerical study of sinusoidal leading edges to control stall onset in axial fan blades at Re=183,000. They carried out the numerical study using OpenFOAM®, an open-source platform, using a RANS cubic implementation of k-$\varepsilon$ closure model. They reported a 30% gain in the lift at the post-stall region for the WHALE4415 airfoil compared to the normal airfoil. Rostamzadeh et al.[50-51] performed computation using unsteady Reynolds-Averaged Navier-Stokes (RANS) model to investigate tubercle effect on the NACA-0021. They utilized the Shear Stress Transport transitional model ($\gamma$-$Re_\theta$) to simulate flow at transition regime (Re = 120,000). The computational investigation explained that the counter-rotating primary and secondary vortices appear over the surface of the airfoil with tubercle at all AOA and the interaction between the vortices in the pre-stall and post-stall region. Skillen et al.[52] applied Large Eddy Simulation (LES) over the same unmodified airfoil (NACA0021) and modified airfoil with wavelength l= 0.11c and A=0.015c (where c is the chord length of the airfoil) at Re =120,000. The LES simulation's grid size was 35 million, and the study focused on the post-stall region (at 20° AOA) only with one AOA. The prediction of lift and drag coefficient was quite close to the measurements for the unmodified airfoil. However, there were significant deviations in the prediction of drag and lift coefficient for the modified case. The leading-edge tubercles act as a passive flow control device that generates the vortex through the leading edge. This generated vortex interacts with the flow separation, thereby delaying the flow separation from the surface. This is the rationale behind the high lift coefficient for the modified case at the post-stall region.

The above discussion strongly suggests the benefits of the leading-edge tubercle over a normal airfoil at the post-stall region. However, none of these investigations discusses the unsteady analysis of the vortex interaction generated from the leading-edge tubercle with flow separation at the pre-stall region. The tubercle effect near the stall angles also needs detailed flow investigation to improve the airfoil performance. Thus, the effect of the vortex generated through tubercle on the laminar separation bubble requires a thorough examination at the pre-stall and post-stall regions. The detailed analysis of tubercle configuration (the amplitude and wavelength variation) is another gap in the open literature to optimize the tubercle configuration. This motivates authors to study the effect of leading-edge tubercle at the transition regime with two different combinations of amplitude and wavelength.

The RANS model can provide the mean flow structure, and it can give satisfactory mean data of aerodynamic properties for both pre and post-stall regions. It is quite well known that the RANS equations often do not yield satisfactory results to predict the complex separated unsteady flows. These flows can be accurately modeled using large-eddy simulation (LES) or direct numerical simulation (DNS) methodologies. However, these alternatives are too expensive concerning the required computer performance[52]. Hence, Hybrid RANS-LES/DES models' application improves to overcome the drawbacks of RANS and LES methods. Usually, in this modeling approach, the RANS types method is used to model the near-wall region while the region away from the wall is modeled using LES. Notably, the RANS and LES region's extent can either be fixed manually or can be determined empirically depending on the problem of interest. Thus, the use of the hybrid models overcome the near-wall computational cost of LES[53-54] and provides unsteady flow field information at the same[55-59].

Since the current study investigates unsteady analysis over the unmodified and modified airfoil at the transition regime, selecting the RANS model for the hybrid RANS/LES depends on the accurate prediction of transition in the flow. Choudhry et al.[60] analyzed the flow around the thick-symmetric NACA 0021 airfoil to understand better the characteristics and effects of long separation bubbles (LoSBs) that exist on such airfoils at low Reynolds numbers. In their article, the authors assessed two recently-developed transition models' predictive capabilities, the correlation-based $\gamma$-$Re_\theta$ model and the laminar-kinetic-energy-based $k_T$-$k_L$-$\omega$ models. The transition prediction capability of the $k_T$-$k_L$-$\omega$ model is superior to the correlation-based $\gamma$-$Re_\theta$ model. So, the $k_T$-$k_L$-$\omega$ model is the best choice for RANS for the hybrid RANS/LES model.

The unsteady analysis over unmodified and modified airfoil provides the qualitative description of the vortex interaction generated through tubercle with the laminar separation bubble. Ribeiro et al.[61] employed the proper orthogonal decomposition (POD) over NACA0012 to identify the coherent structure in the turbulent region. Zhao et al.[62] utilized the POD for NACA 0021 unmodified and modified at transition regime, but the primary focus of the study on the post-stall region. The tubercle as a passive flow control device suppresses the vortex shedding downstream of the airfoil. A proper orthogonal decomposition is employed to quantify the suppression through the energy modes and identify the coherent structures.

In the comprehension of the published literature, it is understandable that the flow separation from airfoil surface near stall is associated with the large-scale vortical structures, which play a significant role in stall characteristics of the airfoil. On the other hand, this is quite important to understand these coherent structures in the interaction between the vortex generated through the leading edge tubercle and laminar separation bubble. Hence, this study investigates the characteristics of these flow structures that would shed light on understanding the flow feature of the modified airfoil, in turn, flow control. The proper orthogonal decomposition (POD) is one of the classical ways that reveal the spatial coherence amongst the structure by extracting the spatial orthogonality. Kosambi[63] first proposed this and later extended it by other researchers like Lumley[64] and Sirovich[65]. Previous works of the present group extensively used this technique for different flow configurations[66-74]. Noticeably, POD, being a powerful tool, identifies the spatial coherence and sheds light on the dynamics of the shear layer towards the effect laminar separation bubble and its control with leading-edge tubercle. Therefore, in the context of the present work, we have utilized 3D POD to characterize the dominant vortical structure of the flow to evaluate the improvement in the aerodynamic performance of the modified airfoil, thereby providing a deeper understanding of the flow control phenomenon.

The paper is organized as follows. Section II presents the mathematical models used for the simulations and grid independence studies. Section III reports the detailed description of the flow structures over the unmodified and modified airfoil, such as the slit width ratio, followed by the conclusions in Section IV.

## II. SIMULATION DETAILS

### A. Governing equations

The governing equations for the conservation of mass and momentum are given by

$$\frac{\partial \bar{U}_i}{\partial x_i} = 0 \tag{1}$$

$$\frac{\partial \bar{U}_i}{\partial t} + \frac{\partial \bar{U}_i \bar{U}_j}{\partial x_j} = -\frac{1}{\rho}\frac{\partial \bar{p}}{\partial x_i} + \nu \frac{\partial^2 \bar{U}_i}{\partial x_i \partial x_j} + \frac{\partial \tau_{ij}}{\partial x_j} \tag{2}$$

Here ¯ represents the Reynolds averaged variables. $\bar{U}_i$ is the fluid velocity, $\bar{p}$ is the fluid pressure, ν is the kinematic viscosity. $\tau_{ij}$ is the Reynolds stress tensor, which is given as follows,

$$\tau_{ij} = \frac{2}{3}k\delta_{ij} - 2\nu_t \bar{S}_{ij} \tag{3}$$

The Reynolds' stress term in the momentum equation is calculated with the help of the transport equations as explained in Appendix - I.

### B. Computational set-up and boundary condition

Figure 1 presents a computational domain of size 30C×30C×0.4285C for three-dimensional simulations. The leading edge of the airfoil represents the origin of the coordinate system. The inlet and outlet boundaries are at the X = −10C upstream and X = 20C downstream, respectively, from the leading edge of the airfoil. The top and bottom boundaries are situated at Y = 15C and Y = −15C from the origin. The side boundary conditions are at Z = 0 and Z = 0.4285C. The lateral distances are far enough to eliminate the far-field boundary effects.

We have used ANSYS® ICEM-CFD[75] to generate the grids used in the current study and carried out a grid-independent test using four different sets of grids. Figure 1 also depicts the grid around an airfoil, where the average value of y+ is found to be less than 1 for all the grids considered herein.

We have performed all the simulations using the open-source CFD toolbox OpenFOAM® [76], where the pressure-velocity coupling is achieved using the PISO algorithm. The discretization of the convection term in the momentum and turbulent variable equations utilizes a bounded second-order scheme, and time integration invokes a second-order backward difference scheme. The algebraic equation for velocity and pressure is solved using the preconditioned conjugate-gradient method. For solving the equations of turbulent variables, the Gauss-Seidel method is employed. Once the initial transient settles, time averaging is performed over ~100 flow-through times ($C/U_\infty$) to compute the statistics, where C is the chord length of airfoil and $U_\infty$ is a uniform inlet velocity.

The inlet boundary utilizes the uniform flow condition, while the outlet boundary employs a non-reflecting convective condition. The top and bottom walls are set as the slip wall, and the no-slip boundary condition (U = $k_L$ = $k_T$ = 0 and a wall-normal zero-gradient condition for ω) is imposed on the unmodified and modified airfoil. As the boundary layer fully resolves up to the viscous layer (the value of $y^+$ lies in the viscous layer of the boundary layer), no wall-function is invoked, and the turbulence variables are directly integrated up to the wall. The turbulence intensity for the present simulation is considered the same as provided in experimental data[77]. At inlets, adequate from the airfoil, the laminar kinetic energy associated with pre-transitional fluctuations is zero.

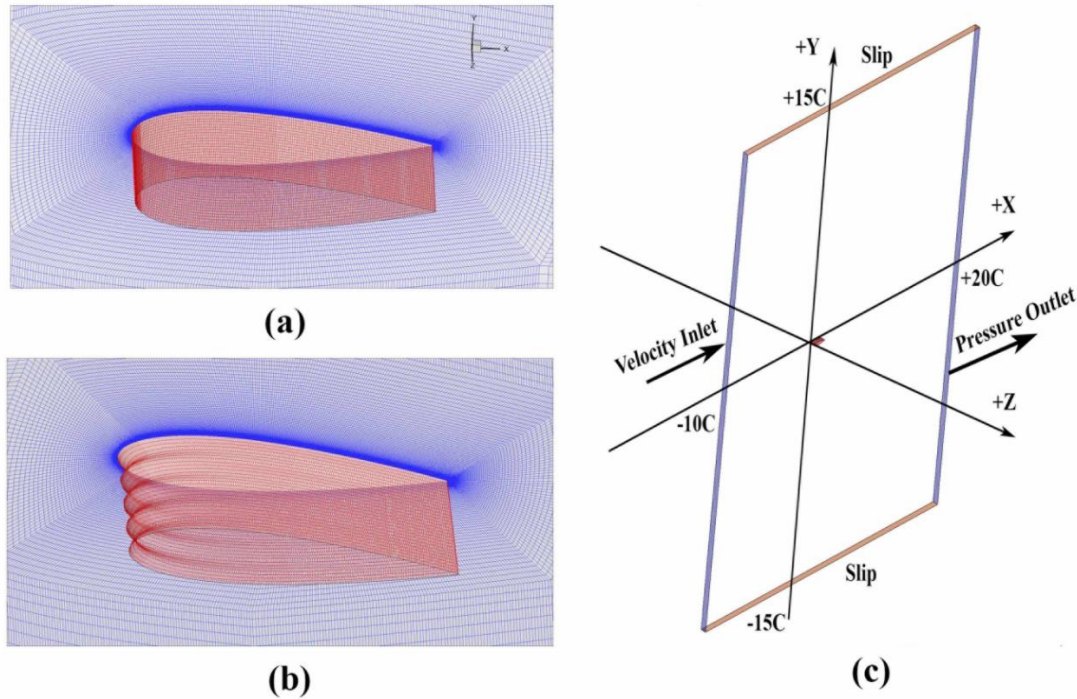

Fig. 1 Computational grid over the (a) unmodified airfoil, (b) modified airfoil, and (c) Schematics of the computational domain and boundary conditions

## C. Grid Independence test

Table I demonstrates the grid independence test performed over the unmodified airfoil (for 20° AOA) with four successive grids. The grid refinement ratio is ~1.50 in each direction over the unmodified airfoil for every consecutive grid from grid-1 to grid-4. The average value of $y^+$ is under one ($y^+<1$) for resolving the boundary layer on the cylinder for each grid. To investigate the sensitivity of the grid independence test, we have considered the mean aerodynamic parameters such as the drag coefficient ($C_D$), the lift coefficient ($C_L$), and the coefficient of pressure ($C_p$). Table I shows the comparison of the $C_L$ and $C_D$ for the different grids, and noticeably, no significant difference exists in the simulated results with grid-3 and grid-4. While comparing the simulated results with the published experimental data, the predictions using grid-3 and grid-4 provide an excellent agreement with the measurements; hence we have chosen grid-3 for the detailed simulations over the airfoil.

TABLE I. Grid independence test for selecting appropriate grid for the simulations

| Grid | Cell No. (million) | $C_D$ | % change ($C_D$) | $C_L$ | % change ($C_L$) |
|---|---|---|---|---|---|
| Grid-1 | 0.3823 | 0.3573 |  | 0.6882 |  |
| Grid-2 | 1.2675 | 0.3338 | 7.0401 | 0.6063 | 13.5081 |
| Grid-3 | 4.2815 | 0.3127 | 6.7476 | 0.5380 | 12.6951 |
| Grid-4 | 14.5046 | 0.3121 | 0.1922 | 0.5295 | 1.6052 |
| Experimental[72] | - | 0.3118 |  | 0.5147 |  |

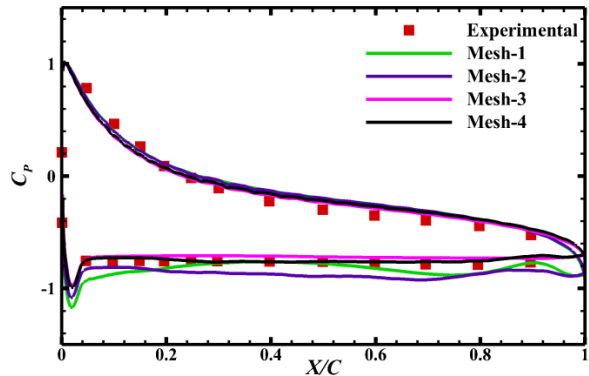

FIG. 2. Coefficient of pressure for different girds compared with the experimental data[77].

The finalized grid for the detailed simulation is also verified with the grid convergence index (GCI). Roache[78] proposed GCI for the uniform reporting of the grid independence studies for numerical simulations in fluids, exploring the theory of generalized Richardson extrapolation to derive the grid refinement error estimator. The detailed description of the grid convergence index (GCI) is found in Ref[78-79]. The terminology of the grids is as follows: grid-2, grid-3, and grid-4 are coarse, medium, and fine grid, respectively. The convergence ratio (R) should be less than one for the monotonic convergence of the system, and it is less than one for the present work, as shown in Table II. The Richardson error estimator for two successive grids is E, and the grid refinement ratio between two consecutive grids is r, and the error ($\varepsilon$) between two grids is evaluated through the discrete solution of two consecutive grids. L2 norms of the grids' errors determine the order of accuracy of the numerical scheme (o). Figure 2 exhibits the L2 norms for the present case and the data compare with the theoretical 2nd order slope. The slope of the L2 norms is also known as the order of accuracy for the numerical schemes, which is at 1.94. Table III provides the GCI for two successive grids. Simultaneously, the value of GCI must reduce for the consecutive grid increment, as estimated using the aerodynamic parameters such as $C_D$ and $C_L$. It can be concluded that the numerical solution shows the reduction in the dependency with the refinement of the grid. The results of GCI also corroborate with the grid independence study (Table I and II), and grid-3 is selected for the investigation over the flow over an airfoil.

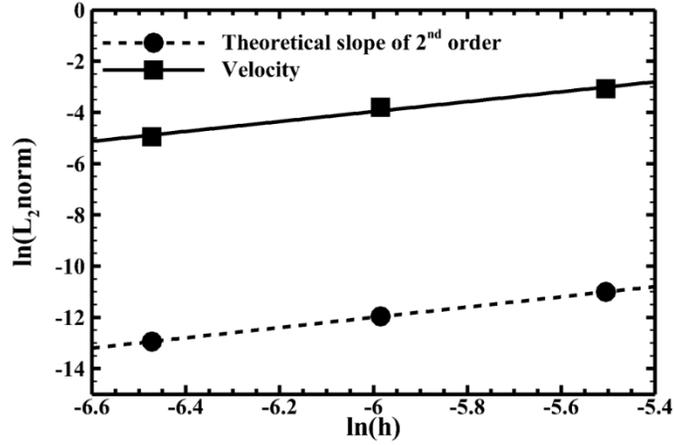

FIG. 3. log-log plot of L$_2$ norm against the grid spacing h

TABLE II. Richardson error estimation and Grid-Convergence Index for three sets of grids

|  | $r_{32}$ | $r_{43}$ | (o) | $\varepsilon_{32}$ ($10^{-2}$) | $\varepsilon_{43}$ ($10^{-2}$) | R | $E^{coarse}$ ($10^{-2}$) | $E^{fine}$ ($10^{-2}$) | $GCI^{coarse}$ (%) | $GCI^{fine}$ (%) |
|---|---|---|---|---|---|---|---|---|---|---|
| $C_D$ | 1.5 | 1.35 | 1.93 | 6.7476 | 0.1922 | 0.0285 | -12.432 | -0.1619 | 15.54 | 0.2023 |
| $C_L$ | 1.5 | 1.35 | 1.93 | 12.6951 | 1.6052 | 0.1264 | -23.389 | -0.3846 | 35.08 | 0.1331 |

## III. RESULTS AND DISCUSSION

This section presents the detailed study performed over the unmodified airfoil and modified airfoil with leading-edge tubercle at transition range (Re = 120,000). The study considers multiple AOA to obtain more profound knowledge about the passive flow control over the airfoil. Mostly, the simulations involve a hybrid RANS/LES model. Simultaneously, two specific cases (at the pre-stall region and post-stall region) also consider the LES model to compare with the hybrid RANS/LES data. The study also utilizes two modified airfoils (as shown in Table III). The variation of the modified airfoil represents the effect of amplitude and wavelength of the leading-edge tubercle on the aerodynamic performance of the airfoil.

TABLE III. Nomenclature of the airfoils for tubercle configuration

| Nomenclature | Amplitude (A) | Wavelength ($\lambda$) |
|---|---|---|
| Unmodified/normal airfoil | 0 | 0 |
| Modified airfoil case-I | 0.01428C | 0.1071C |
| Modified airfoil case-II | 0.05714C | 0.4285C |

### A. Effect of leading-edge tubercle on the lift and drag coefficient

Initially, we report the mean aerodynamic parameters like $C_L$ and $C_D$. Figure 4(a-b) shows the lift and drag coefficient for the normal airfoil, modified airfoil case-I, and modified airfoil case-II at various AOAs. The simulated results of hybrid RANS/LES provide excellent prediction as compared to the experimental data. For the unmodified airfoil, the complicated phenomenon like stall characteristics of the airfoil is also captured accurately. The stall angle of the airfoil is found to be at a = 13°, perfectly matched with the published literature. The hybrid RANS/LES model is also compared with large eddy simulation (LES) for the unmodified airfoil. At 5° AOA, the lift and drag coefficient predictions with LES are similar to the hybrid RANS/LES model. Still, LES results demonstrate the overprediction of the drag coefficient at 20° compared to the hybrid RANS/LES model. The LES requires more refinement in grid resolution to capture accurate aerodynamic property, which may be the reason behind the inadequate prediction.

The computational of the passive flow-controlled modified airfoil case-I produces the closed results compared to the experimental data. Further, one can notice that the leading-edge tubercle alters the stall characteristics of the airfoil, and it shows a very gradual nature of stall behavior. In the post-stall region, the aerodynamics characteristics for modified airfoil exhibit a high lift coefficient with a low cost of drag coefficient. However, no significant improvement is found in aerodynamic properties for modified airfoil than a normal airfoil in the pre-stall region.

Figure 4 also represents the lift and drag coefficient of the modified airfoil with an amplitude A = 0.05714C and the wavelength λ=0.4285C (case-II) for some selected AOAs. The modified airfoil case - II displays poor aerodynamic performance at both pre-stall and post-stall regions compared to modified airfoil case-I. It is interpreted that the aerodynamic properties improve as the amplitude and wavelength decrease. The abated performance of the modified airfoil explains in the following sections.

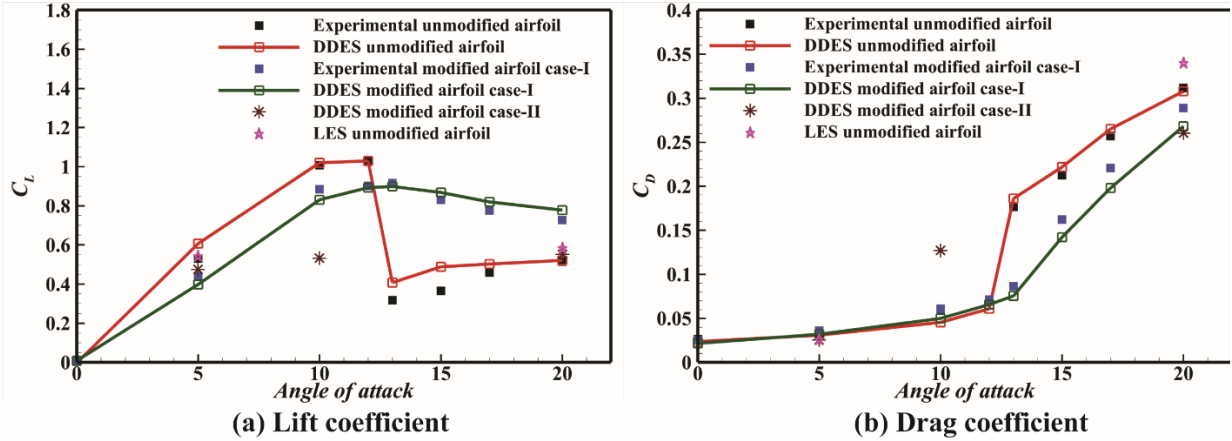

(a) Lift coefficient  (b) Drag coefficient

Fig. 4. (a) Lift and (b) drag coefficient variation with AOA for normal, modified airfoil case-I and modified airfoil case-II. Experimental data is from Hansen[77]

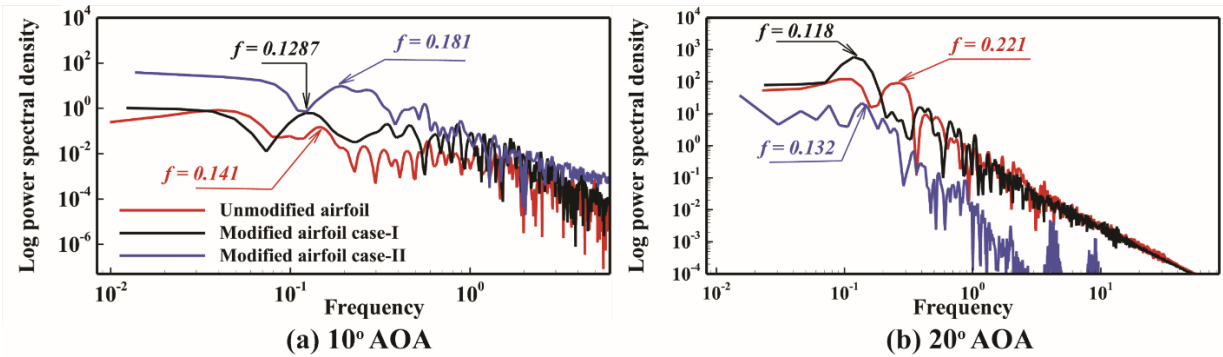

(a) 10° AOA  (b) 20° AOA

FIG. 5. FFT plots of lift coefficients for unmodified, modified cylinder case-I and modified airfoil case-II: (a) 10° AOA, (b) 20° AOA

Figure 5 illustrates the fast Fourier transform (FFT) of the lift coefficient (in the form of power spectral density versus frequency) for the modified and unmodified airfoils at 10° and 20° AOA. At 10° AOA, the FFT plot shows the maximum energy for the modified airfoil case-II, and unmodified airfoil exhibit the least energy. This means the energy of the system increases with the application of the tubercle at the pre-stall region. The lift coefficient's predicted value for the modified case-II at the pre-stall region decreases significantly, and the drag coefficient increases rapidly. Overall, the aerodynamic performance of the modified airfoil case-II reduces as compared to other cases. At 20° AOA,

The FFT plot of modified airfoil case-II illustrates the lowest energy compared to other cases. And the energy for the unmodified airfoil and modified airfoil is almost equal, and this corroborates with the aerodynamic performance of the airfoils. For modified airfoil case-I, the drag coefficient reduces, and the lift coefficient increases compared to the unmodified airfoil. This may be the reason for the approximate same energy content in both cases. The modified case-II demonstrates the lowest lift and drag coefficient, leading to the lowest energy than other cases.

### B.  Effect of leading-edge tubercle on the surface pressure coefficient

Figure 6 presents the coefficient of pressure for unmodified airfoil and modified airfoil case-I at various AOAs. The pressure coefficient ($C_P$) of the normal airfoil displays an excellent comparison with the pre-stall and post-stall region's experimental data. The hybrid RANS/LES model predicts the transition (laminar separation bubble LSB) accurately at the AOA in the pre-stall regime.  We have also performed a large eddy simulation (LES) over the normal airfoil to evaluate and compare the hybrid RANS/LES model with the same selected grid (grid-3). However, LES's lift and drag coefficient prediction is good at the pre-stall and post-stall regions compared to experimental data (Figure 4). However, the LES simulation result significantly lacks the prediction of transition at the pre-stall region (for AOA = 5°), as depicted in Figure 6(b). LES simulation may require more boundary layer resolution or many grid points to predict the transition over the airfoil. This is the primary reason why LES simulation with an unresolved grid cannot capture the laminar separation bubble in the pre-stall regime. However, one must note that the resolved grid LES is well capable of capturing the LSBs. Therefore, we are trying to point out that the performance of the hybrid RANS/LES model is quite impressive at low computational cost while yielding accurate flow physics.

At AOA = 0°, the laminar separation bubbles are of the same size on both the suction and pressure surface of the unmodified airfoil. The laminar separation bubble on the suction surface shifts towards the leading-edge airfoil as we apply the leading-edge tubercle. The modified airfoil case -I also exhibit the laminar separation bubbles on both sides of the airfoil at the mid and trough sections. Hansen[72] has also observed the separation induced transition (LBS) at the trough section for tubercle airfoil with an amplitude A = 0.05714C and the wavelength λ=0.4285C (case-II).  For 10º AOA, the flow shows the transition at the suction surface for modified airfoil case-I, but unlike the unmodified airfoil boundary layer does not re-attach at the suction surface of the air. The reduction in the lift coefficient can correlate to the area under the pressure coefficient curve. Figure 6(b-c) shows that the pressure at the suction surface increases for the modified airfoil case-I, so the lift coefficient of the unmodified should be higher than the modified airfoil case-I.

The lift coefficient plot 4(a) corroborates with the pressure coefficient findings. At 15° AOA, the Cp of modified airfoil case -I is much greater than unmodified airfoil. The lift coefficient plot confirms the stall for the unmodified airfoil. One can notice that the flow over the modified airfoil shows the pre-stall characteristics. The modified airfoil case-I also illustrates the pressure recovery at 20° AOA.

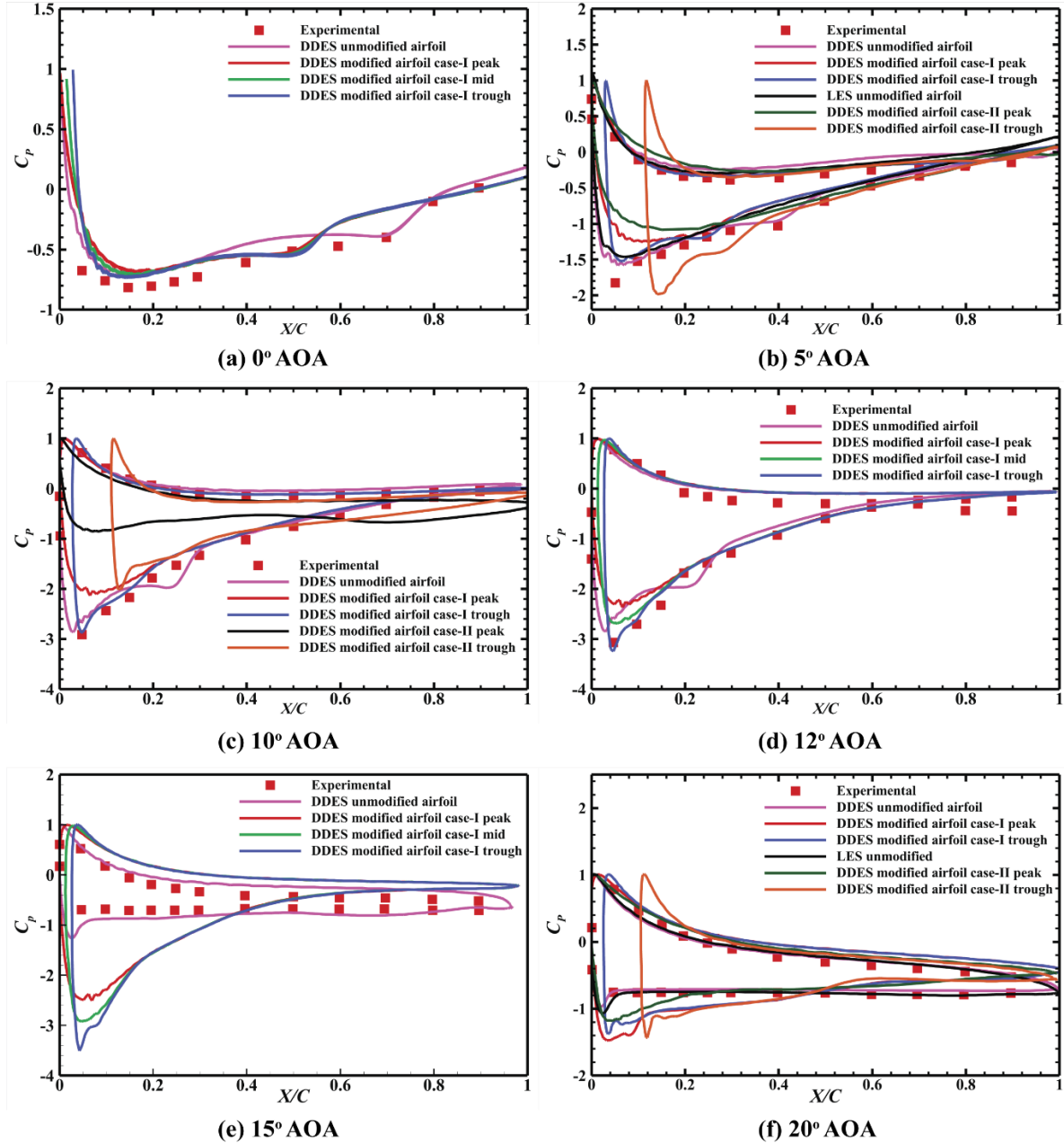

Fig. 6. Coefficient of pressure for normal and modified airfoil at various AOA ($\alpha$). Experimental data is from Hansen[77]

Figures 6 (b, c, and f) also show the coefficient of pressure plots for the modified airfoil case-II. At 5° AOA, the separation-induced transition in the trough section is visible. There is no considerable difference in the area under the pressure coefficient curve, which corroborates the lift and drag coefficient (Figure 4). At 10º AOA, the flow over the peak section shows stall-like behavior, and the pressure coefficient portrays similar behavior as unmodified airfoil at the post-stall region. The sudden rise in drag and abrupt drop in the lift confirm the occurrence stall at the peak section. At the post-stall region (20º AOA), the pressure coefficient plot represents less area than other airfoil cases, leading to the lowest lift coefficient for the modified airfoil case-II. The skin friction coefficient plot is more suitable for representing the laminar separation bubble's length to compare the various cases.

### C.  Effect of leading-edge tubercle on the skin friction coefficient

Figure 7 presents the skin friction coefficient for unmodified and modified airfoil case-I at the suction surface of the airfoil. For modified airfoil case-I, the skin friction coefficient is extracted from three positions, i.e. peak, mid and trough section. Figure 7-a describes the terminology of the separation and re-attachment of flow over the airfoil surface for the unmodified airfoil, and the terminology remains the same for the modified airfoil. At 0º AOA, the laminar separation bubble appears at the trough and mid positions on the modified airfoil case-I. The small separation region is noted on the peak section, which is more clearly visible on the streamlines plots in section III-D. The flow separates at the different locations (X/C) from the modified airfoil case-I suction surface, but the turbulent re-attachment points are almost the same for all (peak, mid, and trough) sections. One can note that the laminar separation bubble length reduces with the application of passive flow control (the leading-edge tubercle) on the unmodified airfoil. At the pre-stall region, the early turbulent separation of flow from the modified airfoil case-I surface (compared to the unmodified airfoil) provides no aerodynamic improvement. The results also corroborate with the lift, drag, and pressure coefficient (Figure 4-5). The laminar separation bubble moves towards the leading edge of the airfoil with the AOA. The stall behavior on unmodified airfoil can be explained by comparing the skin friction coefficient of 12º and 13º AOA. For unmodified airfoil, the flow over the suction surface separates at x/C=0.081 and re-attaches at x/C=0.29 for 12º AOA. However, for 13º AOA, the flow does not re-attach once it separates from the suction surface. The pressure difference between the suction surface and the pressure surface reduces (Figure 6-e), which causes a sudden drop in the lift coefficient (Figure 4-a). This phenomenon is also known as bubble bursting; thus, a rapid fall in lift coefficient occurs at stall angle[80-82]. There is a very subtle difference in the flow separation point for all the sections (peak, mid and trough section) for modified airfoil case-I as we compare the skin friction coefficient of 12º and 13º AOA. The difference in the surface pressure on the suction and pressure side provides no significant difference for 13º AOA, reflecting a high lift coefficient at the stall and post-stall regions in Figure 4-a.

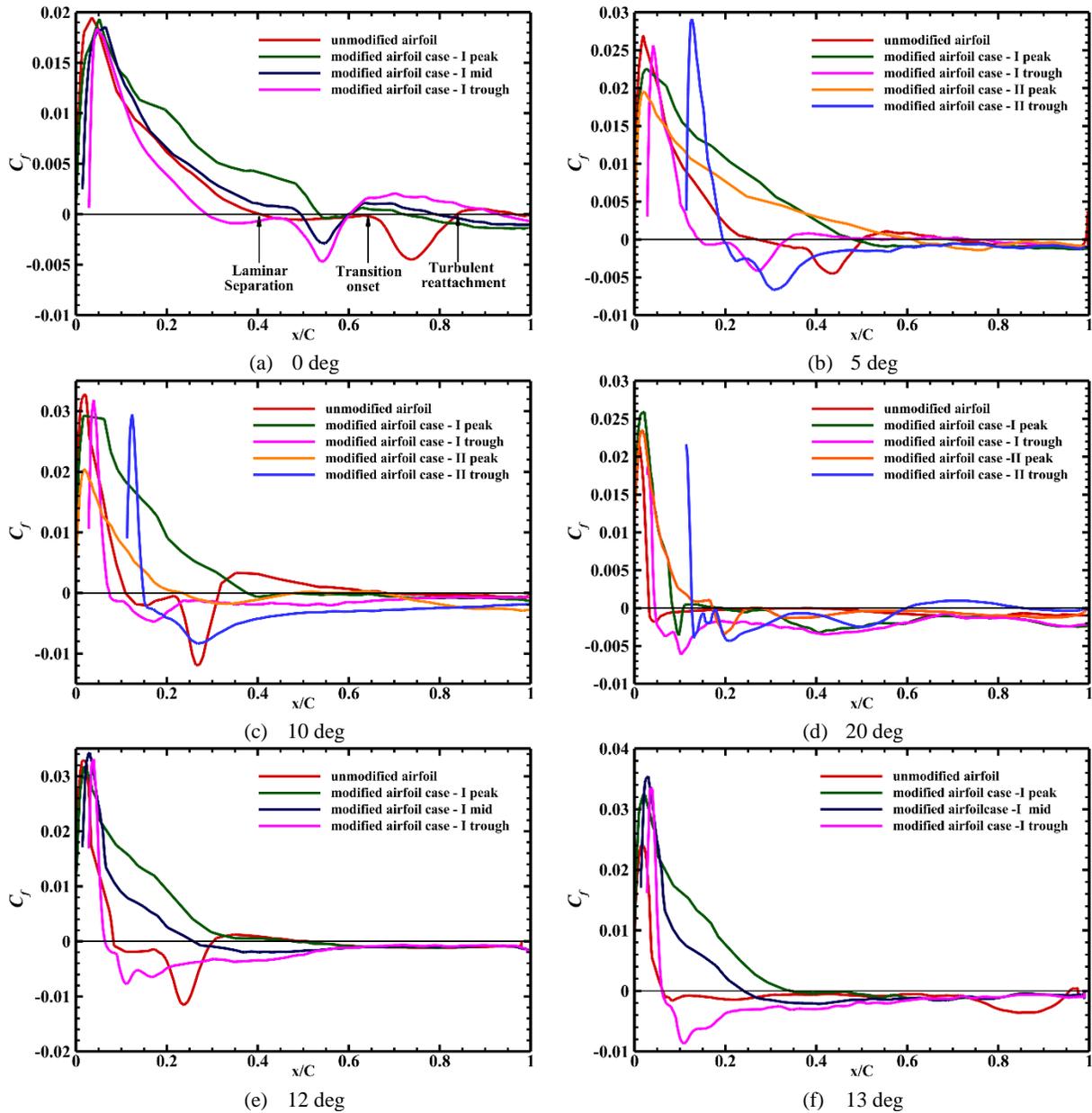

Fig. 7. The skin friction coefficient for normal and modified airfoil at various AOAs: (a) 0 deg, (b) 5 deg, (c) 10 deg, (d) 20 deg, (e) 12 deg, (f) 13 deg

Figures 7(b, c, and d) also show the amplitude and wavelength variation of the modified airfoil (case -II) for a few selected AOAs. At 5º AOA, the increment in the amplitude and wavelength (case-II) delays the separation point at the peak and trough sections. The increase in suction pressure of modified airfoil case -II is more than the modified airfoil

case -I (Figure 6b). The increment in the suction pressure creates a strong spanwise flow (secondary flow) over the airfoil. The secondary flow interacting with the airfoil's primary flow causes the delay in the separation point from the suction surface of the modified airfoil case -II. At 10º AOA, as explained in section (III-B), the flow over the modified airfoil case -II exhibits the stall characteristics. At 20º AOA, the modified airfoil case-II flow has a minute separation delay compared to the modified airfoil case-I. A detailed study of the flow structures is provided in sections III-D and E.

The skin friction coefficient represents the separation and re-attachment of flow over the suction surface of the airfoil, but it lacks to characterize that the flow is in the laminar or turbulent region. The fluctuation of velocity utilizes to find the state of the boundary layer (laminar or turbulent) over the suction surface of the airfoil. Figure 8 illustrates the variation of fluctuation intensity in terms of the maximum root mean square (rms) of u' (x-direction velocity) and v' (y-direction velocity). The fluctuation over the airfoil and downstream of the airfoil is calculated at various x/D. The amplification of disturbance within the shear layer initiates the separation-induced transition at mid-air for flow over airfoil[77]. At the pre-stall region, the exciting thing is to note that the fluctuation in the velocities grows as flow separated from the suction surface, which confirms the existence of laminar flow before the separation of the flow. Afterward, the fluctuation in the velocities increases until flow re-attaches over the suction surface for the normal airfoil. However, for the modified airfoil case -I, the fluctuation intensity grows after re-attachment of flow. The interaction of flows (Primary and secondary flow) on the mid and trough section may cause an early transition, leading to an early rise in velocity fluctuation. The fluctuation v' increases at the downstream of the modified airfoil case -I as compared to the unmodified one. Findings of velocity fluctuation reveal that the leading-edge tubercle as a passive control device grows the downstream fluctuation (explanation is given in section III-E). The tubercle at the airfoil exhibit no aerodynamic improvement at the pre-stall region.

On the other hand, at the post-stall region, the velocity fluctuation over the modified airfoil case -I is higher than the unmodified airfoil. This indicates the spanwise flow interaction creates high disturbance over the airfoil. However, the tubercle plays an essential role in controlling the flow downstream of the airfoil. The flow control can explain as the velocity fluctuation reduces downstream of the modified airfoil. The unmodified flow airfoil separates from the suction surface and creates a big separation zone (compared to modified case-I). This long separation starts shedding from the trailing edge of the airfoil, which constructs a big structure, hence showing high fluctuation for unmodified cases downstream of the airfoil. Section III-E provides the details of the flow structure.

Figures 8 b-c also show the velocity fluctuation for modified airfoil case-II to compare the effect of tubercle configuration variation. At 10º AOA, the flow on the trough section illustrates the localized stall for modified airfoil case-II. The burst laminar separation bubble initiates the local stall at the trough section. As a result, the velocity fluctuation is higher than the modified case-I. For modified case-II, the early flow separation generates the large-scale structure, which produces the rise in velocity fluctuation downstream compared to modified case-I. At 20º AOA, the delay in the separation from the suction surface suppresses the fluctuation for modified case-I.

The above discussion focuses on quantifying the coefficient of pressure variation at the selected section of the airfoil (for modified cases). The following section (III-D) explains the mean flow behavior over the whole span of the airfoil to a deep understanding of the effect of leading-edge tubercle on the laminar separation bubble.

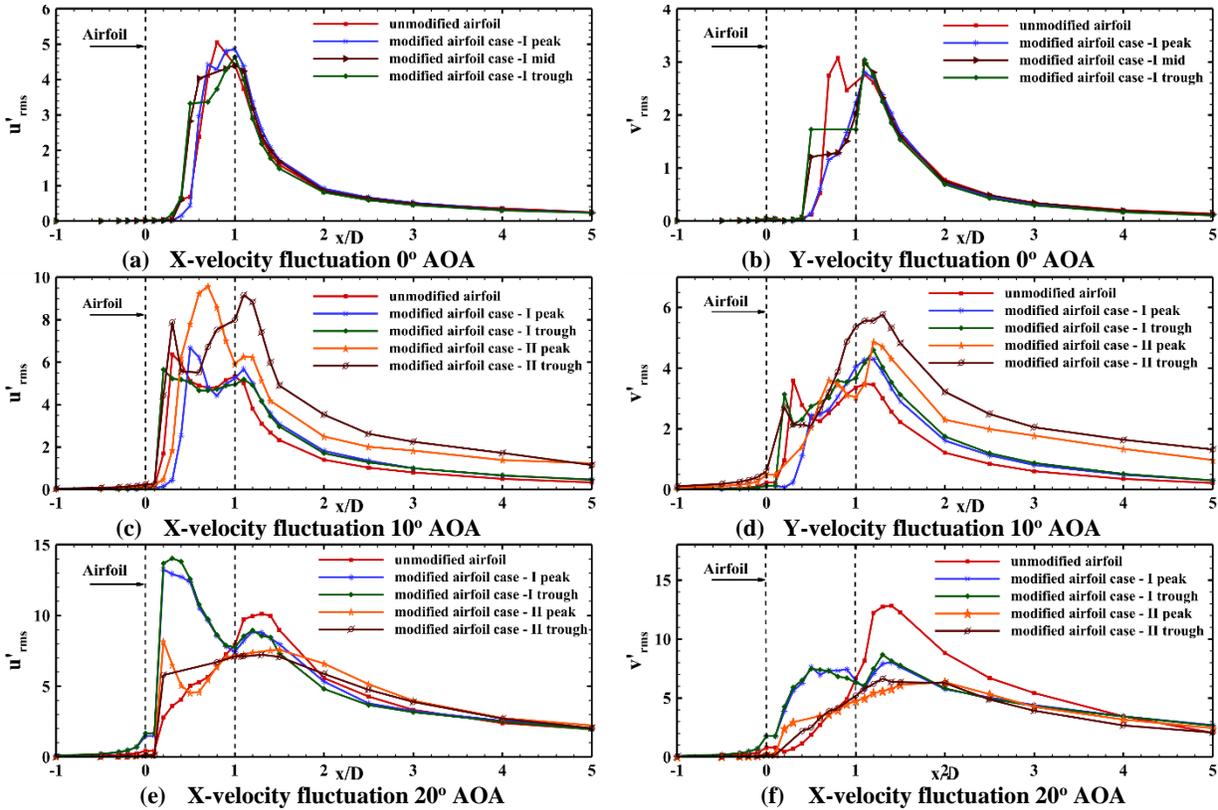

FIG. 8 The maximum root mean square (rms) fluctuation intensities of u′ and v′ along the line of constant x/C at various AOAs: (a) 0º AOA, (b) 0º AOA, (c) 10º AOA, (d) 10º AOA, (e) 20º AOA, (f) 20º AOA

### D. Mean flow analysis for normal and modified airfoil

Figure 9 reports the mean flow streamline plot just above the suction surface of the airfoil at the pre-stall and post-stall regions. The result reveals that the laminar separation bubble shifts towards the leading edge with the AOA, and the laminar separation bubble's size reduce with the AOA. Flow over the normal airfoil represents no span-wise variation over the range of AOA. At 0º AOA, the flow over the airfoil alters with the help of the leading-edge tubercle as a passive control device. The spanwise flow over the modified airfoil case-I interacts with the main flow over an airfoil. Due to flow interaction, the size of the laminar separation bubble reduces compared to the normal cylinder. The direction of secondary flow in the suction surface is from the peak section to the trough section. The boundary layer separation takes place at point P1 from the trough section as a laminar boundary layer. The laminar separation bubble is found at the trough section of the modified airfoil, while the peak section exhibits no laminar separation bubble. The flow interaction on the trough section energizes the flow; thus, the laminar separation bubble size reduction is evident in the trough section compared to the normal airfoil.

The boundary layer separation point (P1) on the trough section moves towards the leading edge of the airfoil with the AOA. At 10º AOA, there is no laminar separation bubble found on the trough section. Once flow separates from the trough, it cannot re-attach to the suction surface. It can be understood that the onset localized stall takes place at the trough section for this AOA. The boundary layer separates from the trough section before the flow separation at normal airfoil, while the separation points on the peak and mid-section delays compared to the normal airfoil. However, the flow re-attaches on the suction surface for normal airfoil leads towards a high lift coefficient for this case compared to modified airfoil (Figure 4). In the post-stall region, the flow over normal airfoil finds early detachment compared to all the modified airfoil sections, which causes the lift gain for the airfoil with the leading-edge tubercle.

The effect of the leading-edge tubercle at the post-stall region appears to be quite significant, and the available literature does not report the same thoroughly. So, we have carried out a detailed investigation at the post-stall region at various AOAs. Figure 10 depicts the mean streamline plots just above the suction surface of the modified airfoil case -I at the post-stall region. The foci point P2 appears to present both sides of the trough section at the suction surface from 0º AOA. The foci point P2 remains situated at every mid-section over the airfoil up to 10º AOA. Due to flow intact between two peaks starts to cross-over the peak and interacts between the mid-sections on both sides of the peak section, some foci point disappears on the mid-section of the airfoil. The interaction of flow over the peak section gets stronger with the AOA. It can be seen from Figures 9 and 10 that the foci point almost disappear for 17º and 20º AOA. It is visible from Figure 11 that the suction surface of the modified airfoil shows the span-wise pressure variation near the leading edge. This pressure gradient is primarily responsible for span-wise flow over the airfoil with the lead edge tubercle. The span-wise pressure gradient increases with the AOA, and it is a favorable pressure gradient for flow. As a result, span-wise velocity (velocity in Z-direction) increases with AOA (Figure 11-b and 11-d). The strong span-wise velocity begins to affect the flow between the two consecutive peaks on the suction surface. This is the reason for disappearing the foci points on the mid-section. The strong velocity may cause a difference in the mean pressure between two consecutive peaks, reflected in 17º and 20º AOA.

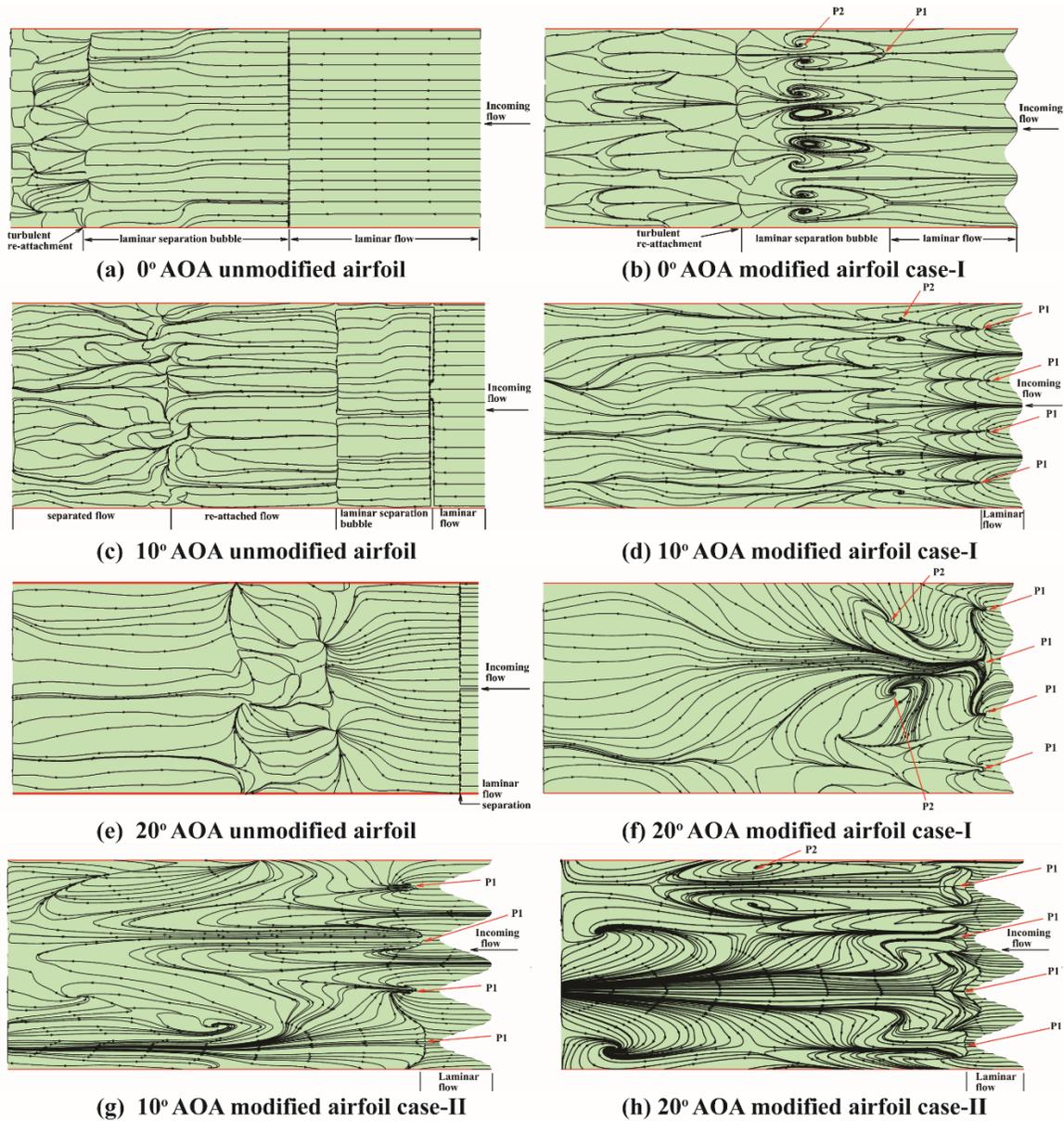

FIG. 9. Mean flow Streamline on the suction side of the airfoil: (a) 0º AOA, (b) 0º AOA, (c) 10º AOA, (d) 10º AOA, (e) 20º AOA, (f) 20º AOA, (g) 10º AOA, (h) 20º AOA

Figure 9 also shows the mean flow streamline plot just above the modified airfoil case-II suction surface at the pre-stall and post-stall regions. At 10º AOA, the streamlines plot exhibits similar characteristics like 17º AOA for the modified airfoil -I. It appears that the strong streamwise flow creates more disturbance, and the flow confined between two peaks starts interacting before 10º AOA in the case of modified airfoil-II. The lowest pressure value increases for 10º AOA in modified airfoil-II (Figure 11-d and 11-g). At 20º AOA, one can notice the difference in the pressure contour plots (Figure 11-f and 11-h). The modified airfoil case-I has more suction pressure than the modified case-II. As a result, modified airfoil-I produces a higher lift than the other two cases.

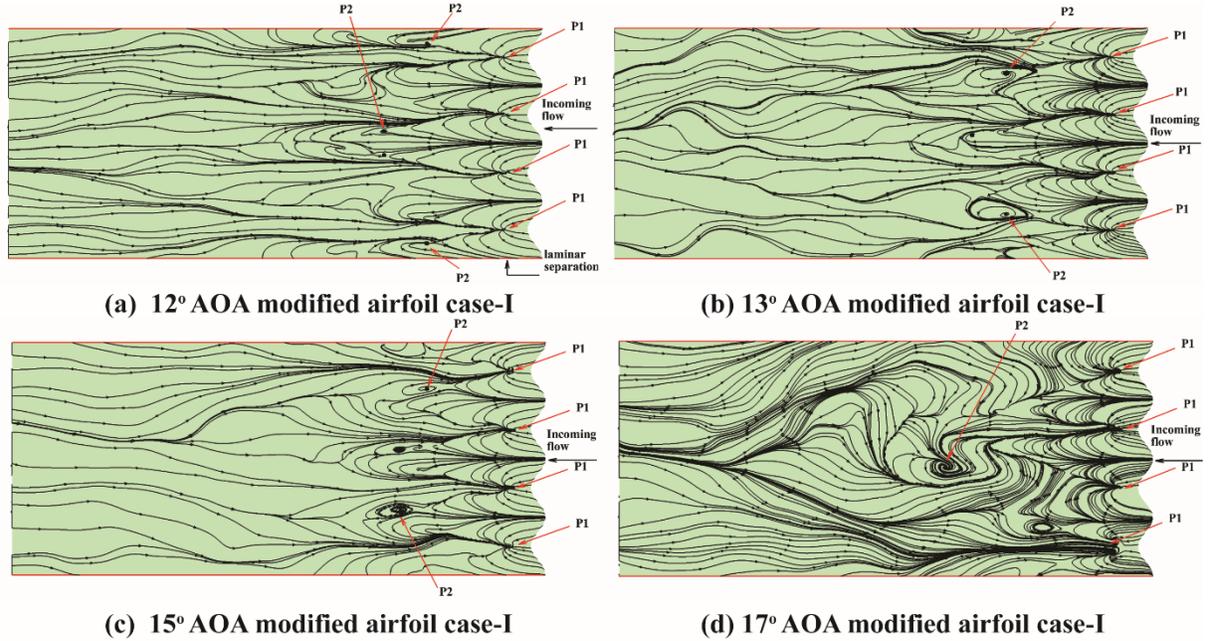

FIG. 10 Mean flow streamlines for the modified case at the post-stall region: (a) 12 deg, (b) 13 deg, (c) 15 deg, (d) 17 deg

This section shows the effect of the mean flow analysis and the pressure contour over the airfoil at the pre-stall and post-stall regions. The instantaneous flow analysis is considered to investigate the flow structure interaction over the airfoil.

**E.  Instantaneous flow analysis**

The instantaneous flow investigation around the airfoil reveals the passive flow control effect of tubercle on NACA0021 airfoil. To corroborate or justify the drag reduction with tubercles, we have considered two AOA, i.e. one is in the pre-stall regime (10º) and one is in the post-stall regime (20º). Figure 12 shows the iso-surfaces of the second invariant of the velocity gradient tensor and instantaneous z-vorticity contour (at peak section) at 10º and 20º AOA. For unmodified airfoil, the flow separates at $X/C = 0.108$ from the normal airfoil's suction surface, then transition occurs at the mid-air, and the boundary layer has enough energy to re-attach as a turbulent boundary layer at $X/C = 0.302$. Finally, the turbulent flow separation occurs from the airfoil surface at $X/C = 0.678$ (Figure 7 c). For modified airfoil, the flow separates at X/C = 0.075 from the trough section of the suction surface, and transition occurs at mid-air, and the flow is not re-attached at the suction surface. The laminar separation occurs from the mid and peak section (on the suction surface) at X/C= 0.3 and 0.38, respectively. Separation-induced transition does not occur at the mid and peak section of the modified airfoil, and these sections exhibit laminar separation.

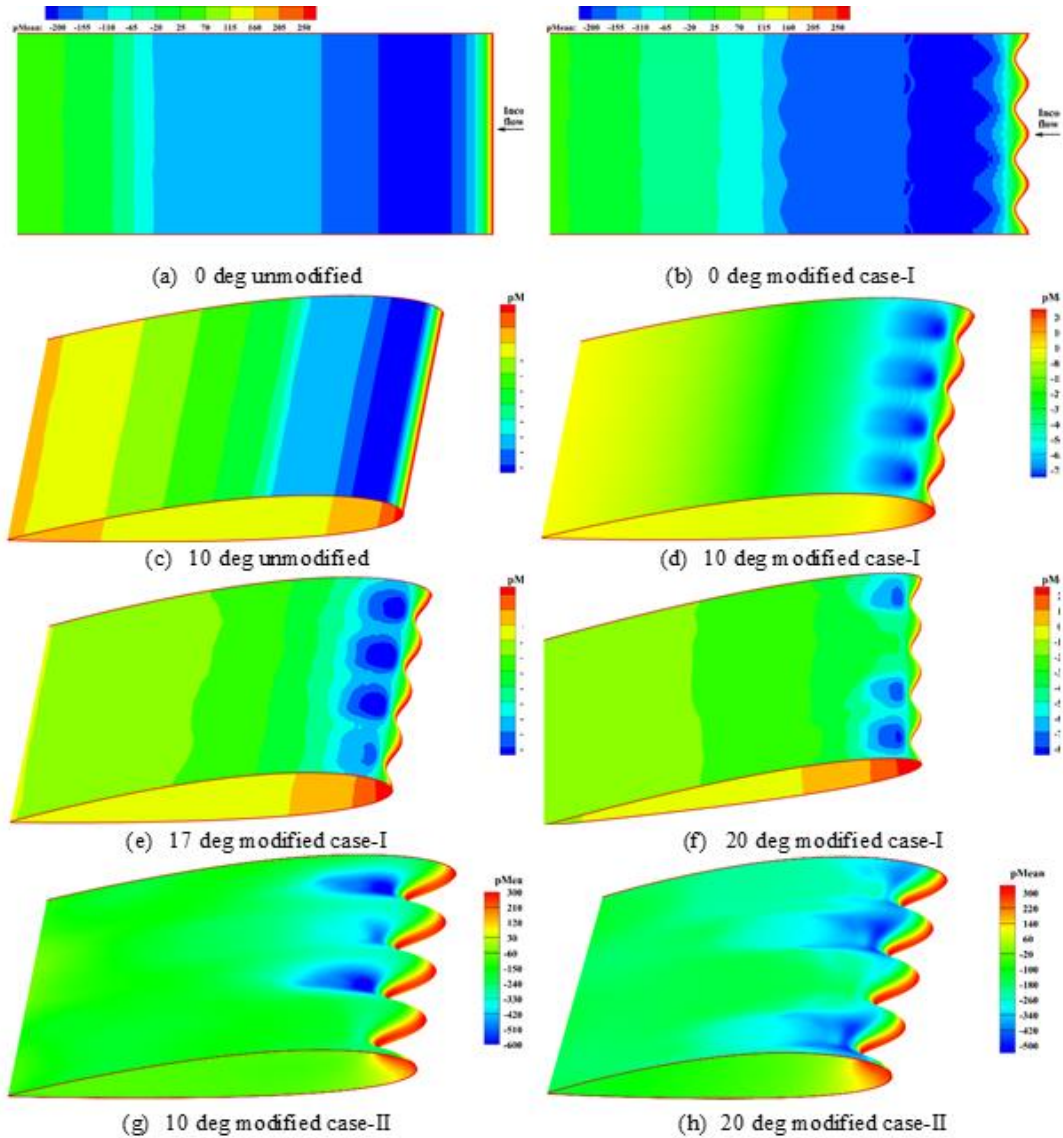

FIG. 11 Mean surface pressure over the unmodified and modified airfoil: (a) 0 deg, (b) 0 deg, (c) 10 deg, (d) 10 deg, (e) 17 deg, (f) 20 deg, (g) 10 deg, (h) 20 deg

**F. Instantaneous flow analysis**

The instantaneous flow investigation around the airfoil reveals the passive flow control effect of tubercle on NACA0021 airfoil. To corroborate or justify the drag reduction with tubercles, we have considered two AOA, i.e. one is in the pre-stall regime (10º) and one is in the post-stall regime (20º). Figure 12 shows the iso-surfaces of the second invariant of the velocity gradient tensor and instantaneous z-vorticity contour (at peak section) at 10º and 20º AOA. For unmodified airfoil, the flow separates at X/C = 0.108 from the normal airfoil's suction surface, then transition occurs at the mid-air, and the boundary layer has enough energy to re-attach as a turbulent boundary layer at X/C= 0.302. Finally, the turbulent flow separation occurs from the airfoil surface at X/C = 0.678 (Figure 7 c). For modified airfoil, the flow separates at X/C = 0.075 from the trough section of the suction surface, and transition occurs at mid-air, and the flow is not re-attached at the suction surface. The laminar separation occurs from the mid and peak section (on the suction surface) at X/C= 0.3 and 0.38, respectively. Separation-induced transition does not occur at the mid and peak section of the modified airfoil, and these sections exhibit laminar separation.

The flow over the suction surface of the unmodified airfoil generates a small structure compared to the modified airfoil-I. The boundary layer becomes turbulent over the entire span of the unmodified airfoil (at X/C = 0.678). This may be the reason for the small flow structure. The flow examination over the modified airfoil-I reveals a fascinating phenomenon over the suction surface of the airfoil. The boundary layer in the trough section separates, and the separated flow confines between two peaks of the modified airfoil case-I. As the separated flow proceeds further downstream, it acquires the shape of the big gap between the two peaks. The structure becomes larger at the trough section and interacts with another large structure from the other trough section. It appears that the overall flow structure from the suction surface of the modified airfoil is larger than the unmodified airfoil. Due to the large structure on the modified airfoil's suction surface, the lift coefficient reduces compared to the unmodified airfoil (Figure 12 a-d). This may be the reason that the leading-edge tubercle airfoil may not produce any significant advantage over a normal airfoil at the pre-stall region.

Figure 13 represents the iso-surfaces of the second invariant of the velocity gradient tensor (Q-criteria) and instantaneous z-vorticity contour (at peak section) at 20º AOA. For unmodified airfoil, the flow separates from the suction surface at X/C = 0.022, and the transition occurs at mid-air. The long re-circulating bubble exists at the suction surface of the unmodified airfoil, causes a drastic drop of the lift for the normal airfoil. For modified airfoil case-I, the flow separates from the trough, mid and peak section (on the suction surface) at X/C = 0.044, 0.058, and 0.083, respectively.

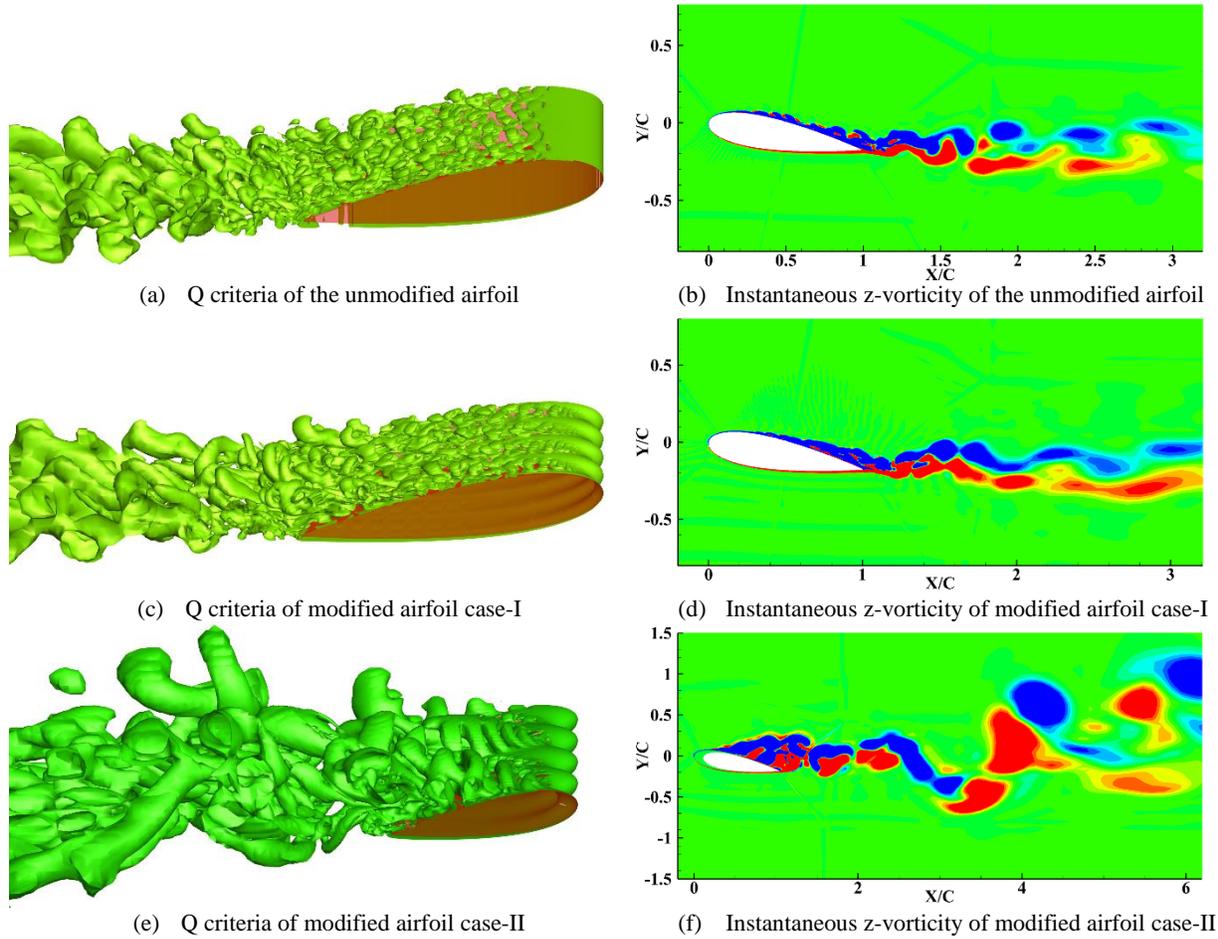

FIG. 12. Iso-surfaces of the second invariant of the velocity gradient tensor, Q = 500 for unmodified and modified airfoils at 10º AOA: (a,c,e) - Q criteria, (b,d,f) – z-vorticity

The delayed separation points for the modified airfoil are the main reason behind the small flow structure compared to an unmodified airfoil. Figures 12 and 13 also represent the iso-surfaces of the second invariant of the velocity gradient tensor (Q-criteria) and instantaneous z-vorticity contour (at peak section) at 10º and 20º AOA for modified airfoil case-II. As explained above, the separated flow from the trough acquires the gap between two peaks. Moreover, the distance between the peaks in modified airfoil case-II is larger than the modified case-I. As a result, the modified airfoil case-II generates the largest structure at 10º AOA. So, the modified airfoil case-II yields poor aerodynamic performance amongst all three airfoils. Similarly, the flow structure over the modified airfoil case-II develops the larges structure than the modified airfoil case-I. However, the overall flow structure is smaller than the unmodified airfoil. It is noticeable that the tubercle on the leading edge of the airfoil significantly improves the performance of the airfoil in the post-stall region.

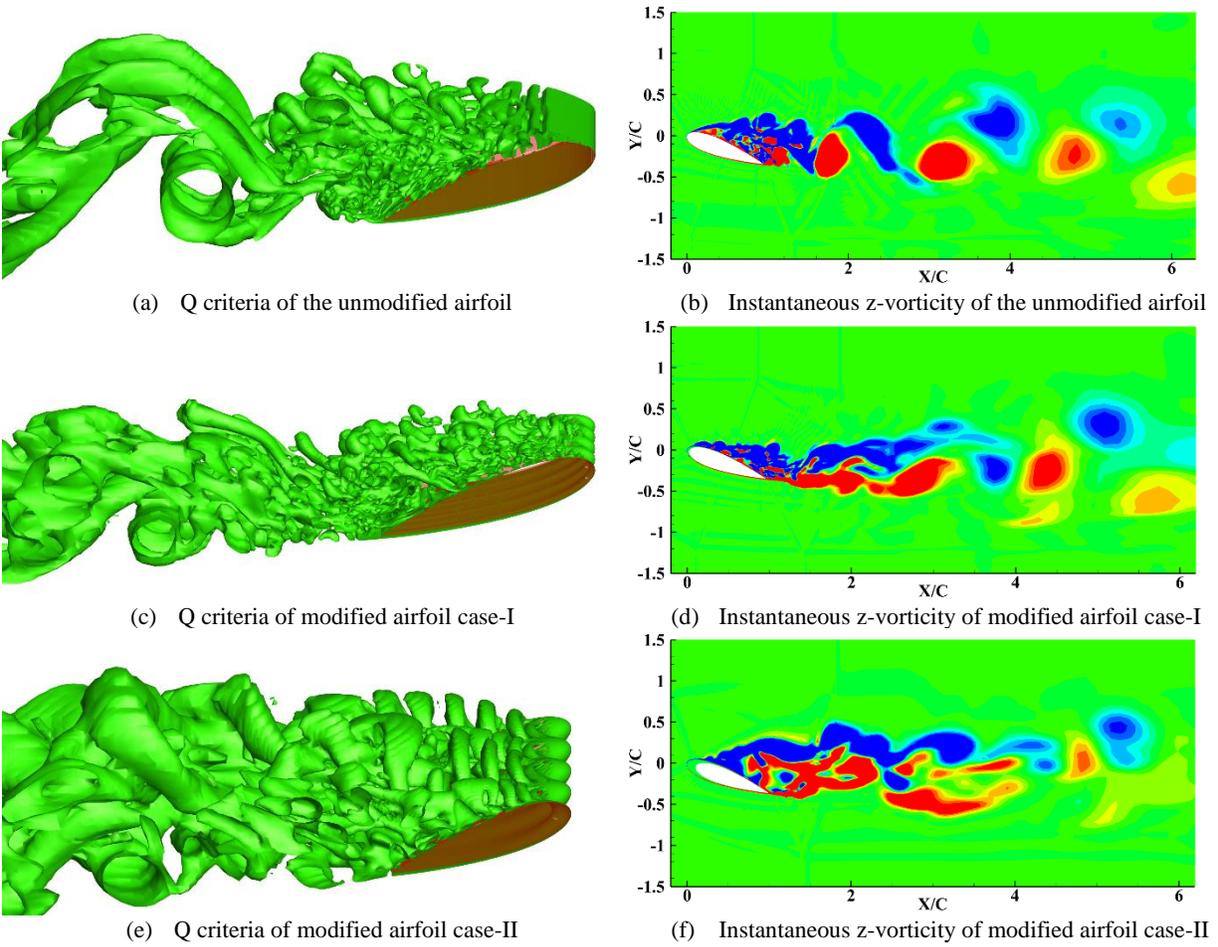

FIG. 13. Iso-surfaces of the second invariant of the velocity gradient tensor, Q = 1000 for unmodified and modified airfoils at 20º AOA: (a,c,e) - Q criteria, (b,d,f) – z-vorticity

The instantaneous investigation provides the significant advantage of leading-edge tubercle on the normal airfoil at the post-stall region. The instantaneous study offers the qualitative comparison of the passive flow-controlled airfoil with the unmodified airfoil. Further, we have employed the proper orthogonal decomposition (POD) method on the instantaneous data to compare the quantity. As explained earlier, the laminar separation bubble on the modified airfoil exhibits a three-dimensional phenomenon. It is essential to apply three-dimension POD analysis for a detailed investigation of the flow over an airfoil.

## G. Proper orthogonal analysis

The proper orthogonal decomposition (POD) technique is a useful mathematical tool to evaluate the dominant flow structure (based on energy contribution). The POD modes based on the respective fluctuation kinetic energy can be calculated through the instantaneous flow field (snapshots), and the extracted POD modes are proportional to their eigenvalues. In most cases, only the first few modes (93-95% cumulative value of energy in modes) are required to represent the features of flow in the system.

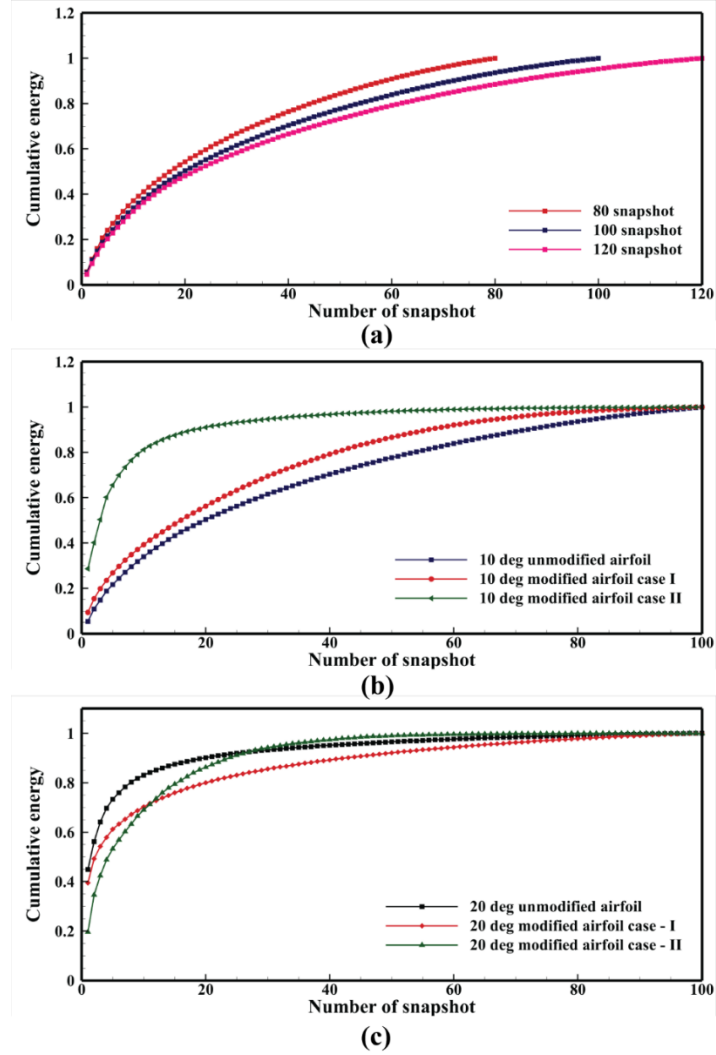

FIG. 14 Cumulative energy plots for (a) the demonstration of snapshots convergence (b) Comparison between modified and unmodified airfoil at 10º AOA (c) Comparison between modified and unmodified airfoil at 20º AOA

The snapshot independence study is performed over three different snapshots to fix the adequate number of snapshots for POD analysis. The finalized grid-3 in section II-B is utilized for the convergence of energy composition with different snapshots. Figure 14 (a) presents the normalized cumulative energy using 80, 100, and 120 snapshots for unmodified airfoil at 10º AOA. The minor over-prediction exits in the first energy mode by the case of 80 snapshots compared to the cases with 100 and 120 snapshots, and the values of modes energy are almost overlapping for 100 and 120 snapshots. The structures of the POD modes show no difference for 80, 100, and 120 snapshots. Hence, 100 snapshots are sufficient enough to carry out the detailed POD analysis for the present study.

Figure 14 (b) provides a comparison between cumulative energy in the pre-stall region. The considered AOA is 10º to analyze the passive flow control at the pre-stall region. The cumulative energy for modified airfoil case-I exhibits more energy to the unmodified airfoil at 10º AOA. The modified airfoil case-II has higher energy modes compared to other cases. This is expected as we explained that the modified airfoil case-II has a larger flow structure than other cases. On the other hand, passive flow control plays a significant role in the post-stall region (Figure 14-c). The modified airfoil case-I is most beneficial to control the long flow separation over the suction surface.

Figure 15 represents the first most energetic mode (mode-1) for unmodified and modified airfoils at the pre-stall region. The iso-surface plots of the second invariant of the velocity gradient tensor and z-direction vorticity clearly show the small flow structure over the suction surface of the unmodified airfoil. Hence, the downstream's flow structure also has a smaller structure compared to the modified airfoils. The relatively larger flow structures exist over the suction surface of modified airfoils due to the confined flow between two peaks attains the large structure (as explained in section III-E). One interesting thing to notice that the vortex structure stretched downstream of the modified airfoil in the x-direction. Due to the leading-edge tubercle, the spanwise flow generates over the suction surface. The x-direction velocity and y-direction velocity of unmodified airfoil distribute in the z-direction due to the tubercle at the leading edge of the airfoil. The strength of z-vorticity reduces for the modified airfoils, and the stretching in the flow structures occurs (Figure 15 b, 15c, and 15e). The results corroborate with the findings of the instantaneous flow investigation (section III-E).

On the other hand, at 20º AOA, the leading-edge tubercle on the airfoil delays the separation point on the suction surface, resulting in a relatively smaller re-circulation zone compared to the unmodified airfoil (Figure 16). As explained in sections III-C and E, the boundary layer separates from the trough section at an almost similar point as the unmodified airfoil, but separation points delay in the mid and peak section. It is observed that the average separation point for the modified airfoil is greater than the unmodified airfoil, which is a primary reason for the better aerodynamic performance in the case of the passive flow-controlled airfoil. One can observe that the flow structures over the modified airfoil illustrate the vortex starching as the pre-stall region. Due to a considerable distance between two peaks, the modified airfoil case -II shows a relatively large ring-type structure, which causes aerodynamic losses compared to the modified airfoil case-II. The modified airfoil-I is the most effective tubercle configuration at both the pre-stall and post-stall regions to improve the aerodynamic performance of the NACA0021 airfoil.

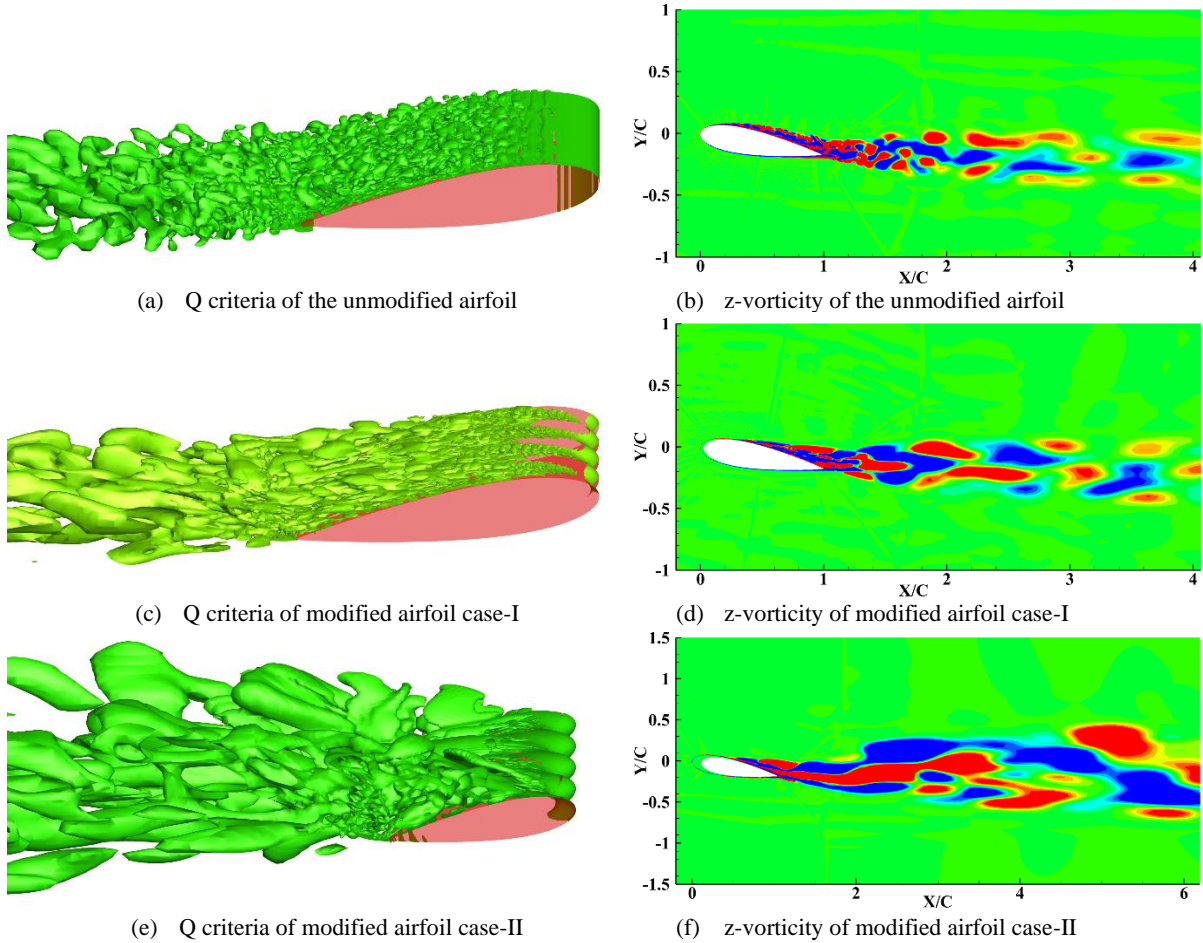

(a) Q criteria of the unmodified airfoil
(b) z-vorticity of the unmodified airfoil
(c) Q criteria of modified airfoil case-I
(d) z-vorticity of modified airfoil case-I
(e) Q criteria of modified airfoil case-II
(f) z-vorticity of modified airfoil case-II

FIG. 15. First most energetic mode in the form of the second invariant of the velocity gradient tensor and z-vorticity of the first mode (at peak section) at 10º AOA: (a,c,e) - Q criteria, (b,d,f) – z-vorticity

## IV. Conclusion

The present study investigates the passive flow control over the airfoil using the leading-edge tubercle as the flow control device. We have invoked a hybrid RANS-LES model *($k_L$-$k_T$-$\omega$ DDES model)* to examine fluid flow behavior over the unmodified and modified airfoils. Further, the validation of the simulated data for flow over an unmodified airfoil (NACA 0021) at 10º AOA shows an excellent agreement with measurements. The grid convergence study provides the validated grid resolution for detailed simulations carried over the range of AOA with the finalized grid. The simulation predicts the stall angle for the unmodified airfoil at a = 13° which is in perfect agreement with the published literature. Further, the simulation shows the excellent prediction of the aerodynamic properties, including the separation-induced transition (size of the laminar separation bubble) for both unmodified and modified airfoil.

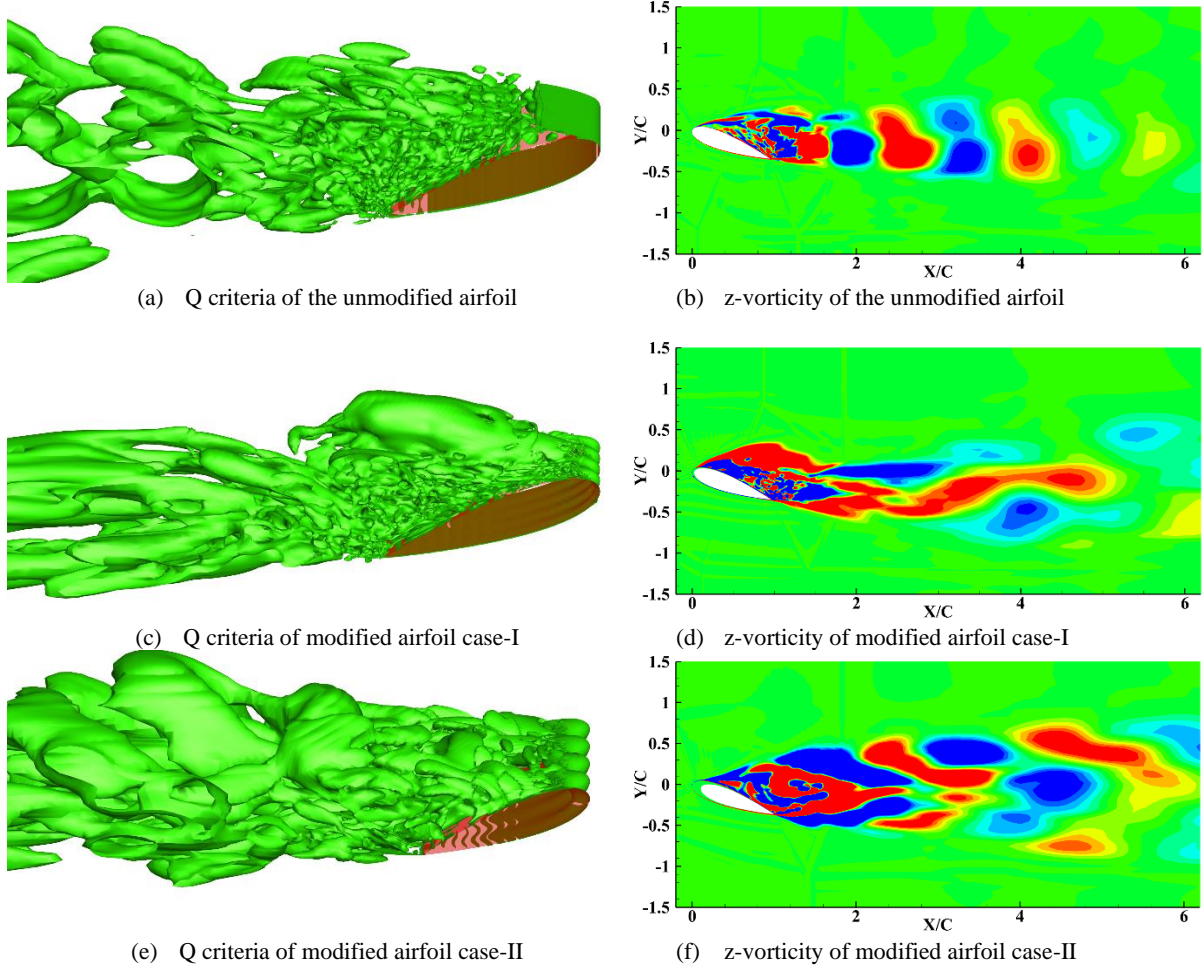

FIG. 16. First most energetic mode in the form of the second invariant of the velocity gradient tensor and z-vorticity of the first mode (peak section) at 20º AOA: (a,c,e) - Q criteria, (b,d,f) – z-vorticity

The leading-edge tubercle in the airfoil creates the spanwise flow due to the geometry of the modified airfoil. The interaction of spanwise flow and main flow over the modified airfoil energizes the flow and delays the average boundary layer separation point, leading to the lift gain for modified airfoil compared to the unmodified airfoil. The skin friction coefficient exhibits a delay in the flow separation on the modified airfoil's suction surface. There is no significant improvement in the aerodynamic properties for modified airfoil at the pre-stall region due to the larger vortical structure present in the modified airfoil's suction surface. The proper orthogonal decomposition (POD) analysis on the unsteady data reveals that the early boundary layer separation occurs at the trough section. The flow structure gains the size between two peak sections. Due to these large structures, there is a loss in the lift coefficient at the pre-stall region for the modified airfoil. The spanwise flow on the modified airfoil causes the vortex structuring in the x-direction. In the post-stall region, the leading-edge tubercle delays the separation, and the vortical structure at the suction surface shows a smaller structure than the unmodified airfoil. It can be concluded that the leading-edge tubercle provides a significant advantage in the post-stall region.

## Data Availability

The data that support the findings of this study are available from the corresponding author upon reasonable request.

## ACKNOWLEDGMENTS

Financial support for this research is provided through the Aeronautical Research and Development Board (ARDB, India. The authors would also like to acknowledge the High-Performance Computing (HPC) Facility at IIT Kanpur (www.iitk.ac.in/cc).

# Appendix – I

## 1. Laminar kinetic energy model ($k_T$-$k_L$-$\omega$)

The $k_T$-$k_L$-$\omega$ transition model is a linear eddy viscosity model based on laminar and turbulent kinetic energy. Three transport equations solve the turbulent kinetic energy ($k_T$), the laminar kinetic energy ($k_L$), and the scale-determining variable ($\omega$) to close the Reynolds stress term in the momentum equation. The omega is defined as $\omega = \varepsilon/k_T$, where $\varepsilon$ is the isotropic dissipation. The three transport equations are described as follows,

$$\frac{\partial k_T}{\partial t} + \bar{u}_j \frac{\partial k_T}{\partial x_j} = P_{k_T} + R_{BP} + R_{NAT} - \omega k_T - D_T + \frac{\partial}{\partial x_j}\left[\left(\nu + \frac{\alpha_T}{\sigma_k}\right)\frac{\partial k_T}{\partial x_j}\right] \quad (4)$$

$$\frac{\partial k_L}{\partial t} + \bar{u}_j \frac{\partial k_L}{\partial x_j} = P_{k_L} - R_{BP} - R_{NAT} - D_L + \frac{\partial}{\partial x_j}\left[\nu \frac{\partial k_L}{\partial x_j}\right] \quad (5)$$

$$\frac{\partial \omega}{\partial t} + \bar{u}_j \frac{\partial \omega}{\partial x_j} = c_{\omega 1}\frac{\omega}{k_T}P_{k_T} + \left(\frac{C_{\omega R}}{f_\omega} - 1\right)\frac{\omega}{k_T}\left(R_{BP} + R_{NAT}\right) - C_{\omega 2}\omega^2 + C_{\omega 3}f_\omega \alpha_T f_\omega^2 \frac{\sqrt{k_T}}{d^3} + \frac{\partial}{\partial x_j}\left[\left(\nu + \frac{\alpha_T}{\sigma_\omega}\right)\frac{\partial \omega}{\partial x_j}\right] \quad (6)$$

The rate of dissipation ($\varepsilon$) in the turbulent kinetic energy equation $k_T$ can be represented as

$$\varepsilon = \omega k_T \quad (7)$$

The effective length scale ($\lambda_{\text{eff}}$) is given by

$$\lambda_{\text{eff}} = \min\left(C_\lambda d, \lambda_T\right) \quad (8)$$

Where $\lambda_T$ is the length scale, given by

$$\lambda_T = \frac{\sqrt{k_T}}{\omega} \quad (9)$$

More details about this transition model can be found in[83].

## 2. One equation eddy viscosity model (LES model)

Smagorinsky[84] proposed the concept of Large Eddy Simulations (LES). In this model, the large-scale turbulent structures are filtered, and the small-scale eddies are modeled. The one equation eddy-viscosity SGS model[85] for LES computations is described as follow:

Here, subgrid stresses $\tau_{ij}$ are modeled in terms of the SGS eddy viscosity as,

$$\tau_{ij} = \frac{2}{3}k_{sgs}\delta_{ij} - 2\nu_t \tilde{S}_{ij} \tag{10}$$

where ~ represents filtering, filter width is taken to be the cube-root of the cell volume and:

The SGS kinetic energy equation is given by

$$\frac{\partial k_{sgs}}{\partial t} + \tilde{u}_j \frac{\partial k_{sgs}}{\partial x_j} = -\tau_{ij}\frac{\partial \tilde{u}}{\partial x_j} - \frac{1}{\Delta}C_\varepsilon k^{\frac{3}{2}}_{sgs} + \frac{\partial}{\partial x_j}\left[\nu_{sgs}\frac{\partial k_{sgs}}{\partial x_j}\right] \tag{11}$$

The eddy viscosity is given by,

$$\nu_{sgs} = C_k \Delta \sqrt{k_{sgs}} \tag{12}$$

The values for the constants $C_k = 0.094$, $C_\varepsilon = 1.048$ and $\Delta$ is the maximum length of the local grid size.

$$\Delta = \max\{\Delta_x, \Delta_y, \Delta_z\} \tag{13}$$

3. **Generalization of Hybrid RANS/LES limiter**

Travin et al.[86] proposed a generalized definition of the detached eddy simulation (DES) limiter, which is given by

$$l_{DES} = \min\{l_{RANS}, l_{LES}\} \tag{14}$$

Where, $l_{RANS}$ and $l_{LES}$ are the RANS length scale and the LES length scale, respectively.

4. **Hybrid RANS/LES model**

The hybrid RANS/LES model (DES) combines $k_T$-$k_L$-$\omega$ transition RANS model and one equation eddy-viscosity SGS model. The DES formulation involves the substitution of the RANS length scale with the DES length scale. The modification is carried out through the turbulent kinetic energy $k_T$ transport equation ($k_{sgs}$ transport equation for the LES model). The length scale $l_{RANS}$ for the RANS model is given by

$$\lambda_T = \frac{\sqrt{k_T}}{\omega} \tag{15}$$

The length scale $l_{LES}$ for the LES model can be represented as

$$l_{LES} \approx C_{DES}\Delta \tag{16}$$

The switching from the $k_T$-$k_L$-$\omega$ RANS model to an SGS model takes place where the turbulent length scale $\lambda_T$ (predicted by RANS model) becomes larger than the local grid size $\Delta$.

The DES length scale $l_{DES}$ is defined as,

$$l_{DES} = \min\{\lambda_T, C_{DES}\Delta\} \tag{17}$$

Applying the DES condition by substituting $\lambda_T$ by $l_{DES}$

$$\omega = \frac{\sqrt{k_T}}{\lambda_T} \quad \text{for} \quad (C_{DES}\Delta > \lambda_T)$$
$$\omega = \frac{\sqrt{k_T}}{(C_{DES}\Delta)} \quad \text{for} \quad (C_{DES}\Delta < \lambda_T) \tag{18}$$

The DES modification is essentially a multiplier to the destruction term in the $k$ transport equation, as illustrated below

$$\frac{\partial k_T}{\partial t} + \bar{u}_j \frac{\partial k_T}{\partial x_j} = P_{k_T} + R_{BP} + R_{NAT} - F_{DES}(\omega k_T) - D_T + \frac{\partial}{\partial x_j}\left[\left(\nu + \frac{\alpha_T}{\sigma_k}\right)\frac{\partial k_T}{\partial x_j}\right] \tag{19}$$

where $F_{DES}$ is given as

$$F_{DES} = \max\left(\frac{\sqrt{k_T}}{C_{DES}\omega\Delta}, 1\right) \tag{20}$$

The DES approach is very advantageous as compared to fully resolved LES due to less resolution requirement. However, the standard DES approach also has severe limitations due to grid dependency and lack of near-wall damping while performing a grid independence test. A refined grid predicts very contradictory and unusual results. To circumvent this problem, the delayed detached-eddy simulation (DDES) is invoked in the present study[87-88].

For DDES modification in the $k$ transport equation is described as

$$\frac{\partial k_T}{\partial t} + \bar{u}_j \frac{\partial k_T}{\partial x_j} = P_{k_T} + R_{BP} + R_{NAT} - F_{DDES}(\omega k_T) - D_T + \frac{\partial}{\partial x_j}\left[\left(\nu + \frac{\alpha_T}{\sigma_k}\right)\frac{\partial k_T}{\partial x_j}\right] \tag{21}$$

Where $F_{DDES}$ is defined as

$$F_{DDES} = \max\left(\frac{\sqrt{k_T}}{C_{DES}\omega\Delta}(1-F_1), 1\right) \tag{22}$$

F1 is the blending function that switches value 0 at the edge of the boundary and F1 = 1 for inside the boundary layer.

$$F_1 = \tanh\left\{\left\{\min\left[\max\left(\frac{\sqrt{k}}{\beta^*\omega y}, \frac{500\nu}{y^2\omega}\right), \frac{4\sigma_{\omega 2}k}{CD_{k\omega}y^2}\right]\right\}^4\right\} \tag{23}$$

More details about the model can be found in ref[89].